# Near-field Fourier ptychography: super-resolution phase retrieval via speckle illumination

**He Zhang,**[1,4,6] **Shaowei Jiang,**[1,6] **Jun Liao,**[1] **Junjing Deng,**[3] **Jian Liu,**[4] **Yongbing Zhang,**[5] **and Guoan Zheng**[1,2,*]

[1]*Biomedical Engineering, University of Connecticut, Storrs, CT, 06269, USA*
[2]*Electrical and Computer Engineering, University of Connecticut, Storrs, CT, 06269, USA*
[3]*Advanced Photon Source, Argonne National Laboratory, Argonne, IL 60439, USA.*
[4]*Ultra-Precision Optoelectronic Instrument Engineering Center, Harbin Institute of Technology, Harbin 150001, China*
[5]*Shenzhen Key Lab of Broadband Network and Multimedia, Graduate School at Shenzhen, Tsinghua University, Shenzhen, 518055, China*
[6]*These authors contributed equally to this work*
*\*guoan.zheng@uconn.edu*

**Abstract:** Achieving high spatial resolution is the goal of many imaging systems. Designing a high-resolution lens with diffraction-limited performance over a large field of view remains a difficult task in imaging system design. On the other hand, creating a complex speckle pattern with wavelength-limited spatial features is effortless and can be implemented via a simple random diffuser. With this observation and inspired by the concept of near-field ptychography, we report a new imaging modality, termed near-field Fourier ptychography, for tackling high-resolution imaging challenges in both microscopic and macroscopic imaging settings. The meaning of 'near-field' is referred to placing the object at a short defocus distance with a large Fresnel number. In our implementations, we project a speckle pattern with fine spatial features on the object instead of directly resolving the spatial features via a high-resolution lens. We then translate the object (or speckle) to different positions and acquire the corresponding images using a low-resolution lens. A ptychographic phase retrieval process is used to recover the complex object, the unknown speckle pattern, and the coherent transfer function at the same time. In a microscopic imaging setup, we use a 0.12 numerical aperture (NA) lens to achieve a NA of 0.85 in the reconstruction process. In a macroscale photographic imaging setup, we achieve ~7-fold resolution gain using a photographic lens. The final achievable resolution is not determined by the collection optics. Instead, it is determined by the feature size of the speckle pattern, similar to our recent demonstration in fluorescence imaging settings (Guo et al., *Biomed. Opt. Express*, 9(1), 2018). The reported imaging modality can be employed in light, coherent X-ray, and transmission electron imaging systems to increase resolution and provide quantitative absorption and phase contrast of the object.

## 1. Introduction

Achieving high spatial resolution is the goal of many imaging systems. Designing a high numerical aperture (NA) lens with diffraction-limited performance over a large field of view remains a difficult task in imaging system design. On the other hand, creating a complex speckle pattern with wavelength-limited spatial features is effortless and can be implemented via a simple random diffuser. In the context of super-resolution imaging, speckle illumination has been used to modulate the high-frequency object information into the low-frequency passband [1-8], similar to the idea of structured illumination microscopy [9-12]. Typical implementations achieve two-fold resolution gain in the linear region. With certain support constraints, 3-fold resolution gain has also been reported [8]. More recently, we have demonstrated the use of a translated unknown speckle pattern [4] to achieve 13-fold resolution gain for fluorescence imaging [13]. Different from previous structured illumination demonstrations, this strategy uses a single dense speckle pattern for resolution improvement and allows us to achieve more than one order of magnitude resolution gain without the direct access to the object plane. Based on the same strategy, we have also improved the resolution of a 0.1-NA objective to the level of 0.4-NA objective [13], allowing us to obtain both large field of view and high resolution for fluorescence microscopy. Along with a different line, speckle illumination has also been demonstrated for phase retrieval and phase imaging. The developments include speckle-field digital holographic microscopy [14], wavefront reconstruction [15], phase retrieval from far-field speckle data [16], super-resolution ptychography [17], near-field ptychography [18-20], among others. In the latter two cases, a translated speckle pattern (i.e., the probe beam) is used to illuminate the complex object and the diffraction measurements are used to recover the object via a phase retrieval process.

Inspired by the concepts of super-resolution ptychography [17] and near-field ptychography [18-20], we explore a new coherent imaging modality in this work. The proposed imaging approach, termed near-field Fourier ptychography, is able to tackle high-resolution imaging challenges in both microscopic and macroscale photographic imaging settings. The meaning of 'near-field' is referred to placing the object at a short defocus distance for converting phase information into intensity variations [21]. Instead of directly resolving fine spatial features via a high-NA lens, we project a speckle pattern with fine spatial features on the object. We then translate the object (or speckle) to different positions and acquire the corresponding intensity images using a low-NA lens. Our microscopic imaging setup is shown in Fig. 1(a)-(b), where we use a scotch tape as the diffuser and project the speckle pattern via a high-NA condenser lens. Based on the captured images, we jointly recover the complex object, the unknown speckle pattern, and the coherent transfer function (CTF) in a phase retrieval process. The final achievable resolution is not determined by the collection optics. Instead, it is determined by the feature size of the speckle pattern, similar to our recent demonstration in fluorescence imaging settings [13]. In this work, we achieved a ~7-fold resolution gain in both a microscopic and a macroscale photographic imaging setting. Figure 1(c)-(d) show the resolution improvement of the near-field FP in a microscope setup.

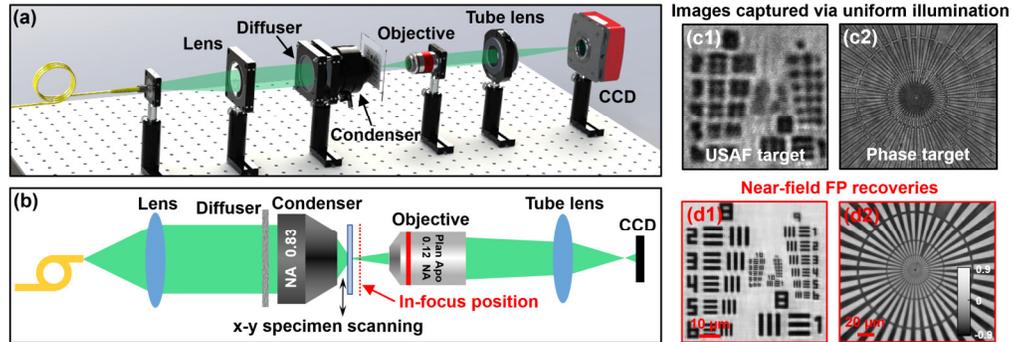

Fig. 1. The near-field Fourier ptychography approach for super-resolution phase retrieval. (a) - (b) The experimental setup, where the object is placed at a short defocus distance to convert complex amplitude into intensity variations. A translated spackle pattern is used for sample illumination and the captured images are used for super-resolution phase retrieval. The captured images of an amplitude (c1) and phase (c2) object under uniform illumination. The recovered intensity of the amplitude object (d1) and the recovered phase of the phase object (d2).

In Fig. 1(b), we place the object at an out-of-focus position in our near-field Fourier ptychography implementation (~40 μm defocused). This short defocus distance is able to convert the phase information of the object into intensity variation in the captured images. The meaning of 'near-field' is, thus, referred to this short distance propagation in our implementation. This strategy is similar to the propagation-based in-line holography which makes transparent objects visible in the intensity measurements [22-24]. It is also similar to the near-field ptychography implementation which has a high Fresnel number [18-20].

The proposed near-field Fourier ptychography is closely related to three imaging modalities: 1) near-field ptychography [18-20], 2) super-resolution ptychography [17], and 3) Fourier ptychography (FP) [25-35]. Drawing connections and distinctions between the proposed approach and its related imaging modalities helps to clarify its operation. Near-field ptychography uses a translated speckle pattern to illuminate the object over the entire field of view. It then jointly recovers both the complex object and the speckle pattern in a ptychographic phase-retrieval process. Similar to the near-field ptychography, the near-field FP setup also captures multiple images under a translated speckle illumination. Since near-field FP is implemented based on a lens system, it appears as the Fourier counterpart of near-field ptychography, justifying the proposed name. By implementing it via a lens system, near-field FP is able to bypass the resolution limit of the employed objective lens.

Super-resolution ptychography uses speckle illumination to improve the achievable resolution. With super-resolution ptychography, the illumination probe is confined to a limited region in the object space, leading to a large number of image acquisitions. With near-field FP, we illuminate the object over the entire field of view. The use of a lens system also enables its implementation in a macroscale photographic imaging setting which may be impossible for regular ptychography approaches.

FP illuminates the object with angle-varied illumination and recovers the complex object profile. Near-field FP replaces the angle-varied illumination with a translated speckle pattern. Both FP and near-field FP can be implemented in a macroscale photographic imaging setting [26, 29]. FP, however, cannot be used for fluorescence imaging because the captured fluorescence images under angle-varied illumination remain identical. The near-field FP setup, on the other hand, is able to improve the resolution of fluorescence microscopy thanks to the use of non-uniform speckle patterns [4, 13]. As such, it can be used in both coherent and incoherent imaging settings to improve the achievable resolution.

In coherent X-ray imaging, it is relatively challenging to generate angle-varied illumination to implement FP [36]. Near-field FP, on the other hand, can be implemented in a regular zone-plate-based transmission X-ray microscope platform without hardware modification. It also allows zone-plate optics to be replaced by total-reflection mirrors [37] or Kinoform diffractive lenses [38], even they are limited to lower NAs or with introduced aberrations. The reported near-field FP may be able to improve the imaging performance of current synchrotron beamline setups and table-top transmission X-ray microscopes.

This paper is structured as follows: in Section 2, we discuss the forward modeling and recovery procedures of the reported approach. Section 3 reports our experimental results on a microscopic imaging setup, where we use a 0.12-NA lens to achieve a NA of 0.77 in the reconstruction process. Section 4 reports our experimental results on a macroscale photographic imaging setup, where we achieve 7-fold resolution gain using a photographic lens. Finally, we summarize the results and discuss future directions in Section 5. The open-source code is provided in the Appendix.

## 2. Modeling and simulations

The forward imaging model of the captured image in near-field FP can be described as

$$I_j(x,y) = \left|\left(O(x - x_j, y - y_j) \cdot P(x,y)\right) * PSF(x,y)\right|^2 \qquad (1)$$

where $I_j(x,y)$ is the $j^{th}$ intensity measurement ($j = 1,2,3…, J$), $O(x,y)$ is the complex object, $P(x,y)$ is the unknown speckle pattern, $(x_j, y_j)$ is the $j^{th}$ positional shift of the specimen (or the speckle pattern), $PSF(x,y)$ is the point spread function (PSF) of the imaging system, and '*' stands for the convolution operation.

Based on all captured images $I_j(x,y)$, the goal of the near-field FP is to recover the complex $O(x,y)$, $P(x,y)$, and $PSF(x,y)$. We have already demonstrated a reconstruction scheme for positive $O(x,y)$ and $P(x,y)$ in an incoherent imaging setting [13]. In the current framework, $O(x, y)$ and $P(x, y)$ are complex values and a phase retrieval process is needed for recovery. The proposed recovery process is outlined in Figs. 2-3, where the captured intensity images are addressed in a random sequence, $O_j(x, y)$ and $P_j(x, y)$ represent the updated object and speckle using the $j^{\text{th}}$ captured image. In this process, we initialize the amplitude of the object by shifting back the captured raw images and averaging them. We then initialize the amplitude of the unknown speckle pattern by averaging all measurements. Finally, we initialize the PSF using the pre-set defocus distance.

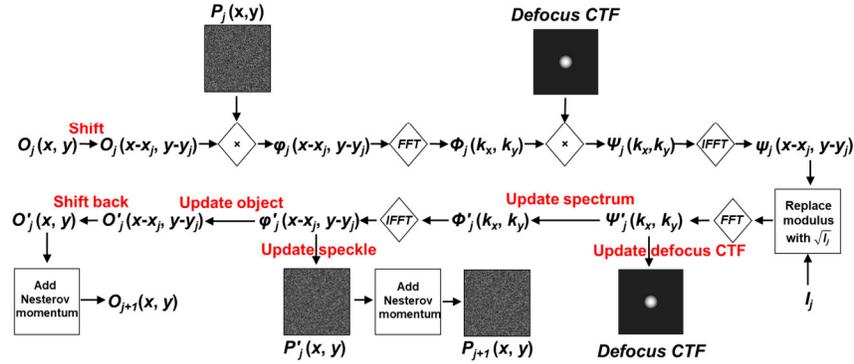

Fig. 2. The flowchart of the recovery process.

---

**Algorithm outline**

---

**Input**: Raw image sequence $I_j$ ($j = 1, 2, \cdots, J$)
**Output**: High-resolution object $O(x, y)$ and unknown speckle pattern $P(x, y)$.

---

Initialize $O(x, y)$, CTF, and $P(x, y)$, and let $O_1(x, y) = O(x, y)$, $P_1(x, y) = P(x, y)$, $t = 0$
**for** $n = 1: N$ (different iteration loops)
    **for** $j = 1: J$ (different captured images)
       $t = t+1$ (for adding Nesterov momentum)
       $O_j(x - x_j, y - y_j) = \text{Shift}\left(O_j(x, y)\right)$
       $\varphi_j(x - x_j, y - y_j) = O_j(x - x_j, y - y_j) \cdot P_j(x, y)$
       $\Phi_j(k_x, k_y) = FFT\left(\varphi_j(x - x_j, y - y_j)\right)$
       $\Psi_j(k_x, k_y) = \Phi_j(k_x, k_y) \cdot CTF$
       $\psi_j(x - x_j, y - y_j) = IFFT\left(\Psi_j(k_x, k_y)\right)$
       $\Psi'_j(k_x, k_y) = FFT\left(\psi_j(x - x_j, y - y_j) / |\psi_j(x - x_j, y - y_j)| \cdot \sqrt{I_j}\right)$
       Update spectrum: $\Phi'_j(k_x, k_y) = \Phi_j(k_x, k_y) + \text{conj}(CTF) \cdot \left(\Psi'_j(k_x, k_y) - \Psi_j(k_x, k_y)\right) / \max(|CTF|^2)$
       Update $CTF$: $CTF = CTF + \gamma_{CTF} \cdot \text{conj}(\Phi'_j(k_x, k_y)) \cdot \left(\Psi'_j(k_x, k_y) - \Psi_j(k_x, k_y)\right) / \max(|\Phi'_j(k_x, k_y)|^2)$
       $\varphi'_j(x - x_j, y - y_j) = IFFT\left(\Phi'_j(k_x, k_y)\right)$
       Update object: $O'_j(x - x_j, y - y_j) = O_j(x - x_j, y - y_j) +$
           $\gamma_{obj} \cdot \text{conj}(P_j(x, y)) \cdot \left(\varphi'_j(x - x_j, y - y_j) - \varphi_j(x - x_j, y - y_j)\right) / \left((1 - \alpha_{obj}) \cdot |P_j(x, y)|^2 + \alpha_{obj} \cdot \max(|P_j(x, y)|^2)\right)$
       Update pattern: $P'_j(x, y) = P_j(x, y) + \gamma_P \cdot \text{conj}(O'_j(x - x_j, y - y_j)) \cdot \left(\varphi'_j(x - x_j, y - y_j) - \varphi_j(x - x_j, y - y_j)\right) / \left((1 - \alpha_P) \cdot |O'_j(x - x_j, y - y_j)|^2 + \alpha_P \cdot \max(|O'_j(x - x_j, y - y_j)|^2)\right)$
       $O'_j(x, y) = \text{Shift Back}\left(O'_j(x - x_j, y - y_j)\right)$
       If $t == T$
           $t = 0$; add Nesterov momentum:
           $v_{obj,j} = \eta_{obj} \cdot v_{obj,j-T} + \left(O'_j(x, y) - O_{(j+1-T)}(x, y)\right)$ and $O_{(j+1)}(x, y) = O'_j(x, y) + \eta_{obj} \cdot v_{obj,j}$
           $v_{P,j} = \eta_P \cdot v_{P,j-T} + \left(P'_j(x, y) - P_{(j+1-T)}(x, y)\right)$ and $P_{(j+1)}(x, y) = P'_j(x, y) + \eta_P \cdot v_{P,j}$
       end
    end
end

---

Fig. 3. The outline of the recovery algorithm.

In the reconstruction process (Figs. 2-3), we first shift the updated object $O_j(x, y)$ by $(x_j, y_j)$ in the $j^{\text{th}}$ iteration, and multiply with the speckle pattern $P_j(x, y)$ to form an exit wave $\varphi_j(x - x_j, y - y_j)$. We then perform a Fourier transform of $\varphi_j(x - x_j, y - y_j)$ to obtain $\Phi_j(k_x, k_y)$ in the Fourier space. The resulting $\Phi_j(k_x, k_y)$ is multiplied with the defocus coherent transfer function (CTF) to obtain $\Psi_j(k_x, k_y)$. We then perform an inverse Fourier transform of $\Psi_j(k_x, k_y)$ to obtain the complex amplitude $\psi_j(x - x_j, y - y_j)$ on the image plane and perform Fourier magnitude projection to obtain $\Psi'_j(k_x, k_y)$. We use the extended ptychographical iterative engine (ePIE) [39] to update the high-resolution Fourier spectrum $\Phi'_j(k_x, k_y)$ and the defocus CTF in the Fourier domain:

$$\Phi'_j(k_x, k_y) = \Phi_j(k_x, k_y) + \gamma_\Phi \cdot \frac{\text{conj}(CTF) \cdot (\Psi'_j(k_x, k_y) - \Psi_j(k_x, k_y))}{\max(|CTF|^2)}, \quad (2)$$

$$CTF = CTF + \gamma_{CTF} \cdot \frac{\text{conj}(\Phi'_j(k_x, k_y)) \cdot (\Psi'_j(k_x, k_y) - \Psi_j(k_x, k_y))}{\max(|\Phi'_j(k_x, k_y)|^2)}, \quad (3)$$

Based on the updated $\Phi'_j(k_x, k_y)$, we perform an inverse Fourier transform to obtain $\varphi'_j(x - x_j, y - y_j)$. We then use the regularized ptychographical iterative engine (rPIE) [40] to update the object and the speckle pattern in the spatial domain:

$$O'_j(x - x_j, y - y_j) = O_j(x - x_j, y - y_j) + \frac{\text{conj}(P_j(x,y)) \cdot (\varphi'_j(x-x_j,y-y_j) - \varphi_j(x-x_j,y-y_j))}{\left((1-\alpha_{obj}) \cdot |P_j(x,y)|^2 + \alpha_{obj} \cdot \max(|P_j(x,y)|^2)\right)}, \quad (4)$$

$$P'_j(x, y) = P_j(x, y) + \frac{\text{conj}(O'_j(x-x_j,y-y_j)) \cdot (\varphi'_j(x-x_j,y-y_j) - \varphi_j(x-x_j,y-y_j))}{\left((1-\alpha_P) \cdot |O'_j(x-x_j,y-y_j)|^2 + \alpha_P \cdot \max(|O'_j(x-x_j,y-y_j)|^2)\right)}, \quad (5)$$

where $\alpha_{obj}$ and $\alpha_P$ are algorithm weights in rPIE. We add Nesterov momentum to accelerate the convergence speed in our implementation (typically 2-3 times faster for convergence). The detailed implementation and parameter choice are provided in the Appendix (Figs. 11-14).

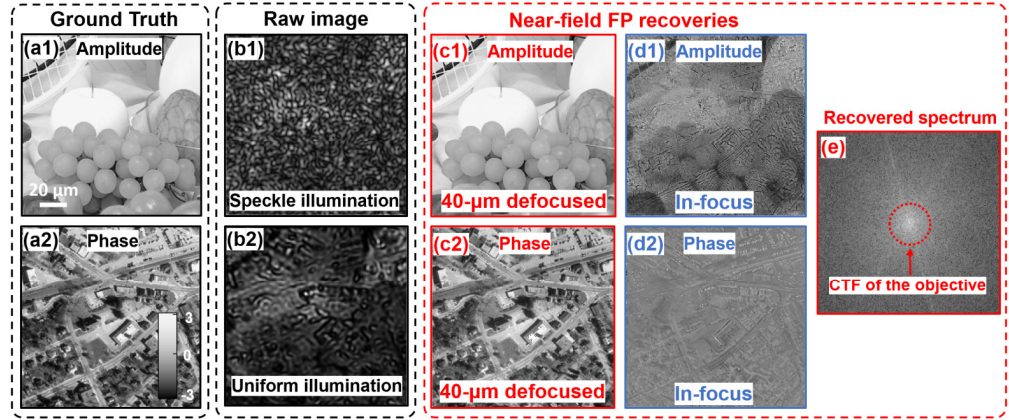

Fig. 4. Simulation of the near-field FP for super-resolution phase retrieval. (a) The ground-truth object. (b) The simulated raw image under speckle and uniform illumination. Near-field FP recoveries by placing the object at a 40-μm defocus distance (c) and in-focus position (d). (e) The recovered Fourier spectrum for (c), where the circle indicates the original CTF of the objective lens.

We have performed several simulations to validate the proposed approach. Figure 4 shows the simulation results that demonstrate the super-resolution phase retrieval concept. In this simulation, the object ground truth is shown in Fig. 4(a). The simulated captured images under speckle and uniform illumination are shown in Fig. 4(b). Figure 4(c) shows the recovered results by placing the object at the 40-μm defocus position. As a comparison, Fig. 4(d) shows the recovered results for the in-focus position. We can see that the amplitude and phase are well-resolved for a short defocus distance, justifying the 'near-field' concept of the proposed approach. The Fourier spectrum of the recovered complex object is shown in Fig. 4(e), where the red dashed circle indicates the original CTF boundary of a 0.12-NA objective lens. The resolution improvement is significant as the recovered Fourier spectrum occupies the entire Fourier space in Fig. 4(e). The computational time is ~6 mins for processing 1681 images with 512 pixels by 512 pixels each using a Dell XPS 8930 desktop computer.

In the second simulation, we analyze the performance of the reported approach with different numbers of translated positions and different noise levels. Figures 5(a)-5(b) show the recovered complex object using different numbers of translated positions. In Fig. 5(c), we quantify the result using root mean square (RMS) error. Based on Figs. 5(a)-5(c), we can see that more translated positions lead to an improved reconstruction and an accelerated convergence. In Figs. 5(d)-5(f), we simulate and analyze the effect of additive noise. Figures 5(d)-5(e) show the recovered object with different amounts of Gaussian noises added into the raw images. The noise performance is quantified in Fig. 5(f).

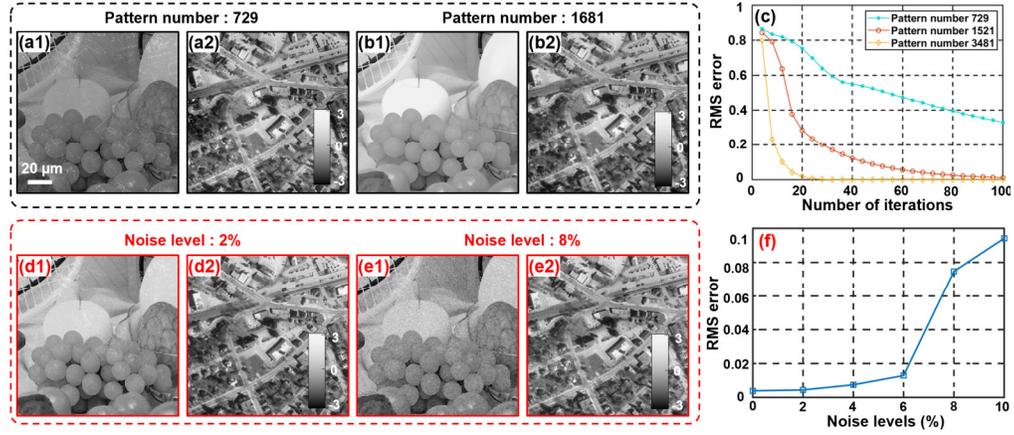

Fig. 5. The imaging performance with different numbers of translated positions and different noise levels. (a)-(b) The recovered object with different numbers of translated positions. (c) The RMS errors are plotted as a function of iterations. (d)-(e) The recovered object with different noise levels; the performance is quantified via the RMS error in (f).

## 3. Near-field FP for microscopy imaging

The experimental setup for microscopic imaging is shown in Fig. 1(a), where we use a 200-mW 532-nm laser diode coupled with a single-mode fiber as the light source. A scotch tape (diffuser) is placed at the back focal plane of a 0.83-NA condenser lens to generate dense speckle pattern on the object. Since the speckle pattern propagates along the axial direction, no focusing of the condenser lens is needed. At the detection path, we use a low-NA objective lens (4X, 0.12 NA) and a monochromatic camera (Pointgrey BFS-U3-200S6) to acquire the object images with a ~0.1 ms exposure time. Two motorized stages (ASI LX-4000) are used to move the object to different x-y positions with 0.25-0.75 μm different and known step size. The acquisition time for ~1000 images is ~3 mins in our experiment. The captured images are then used to recover the complex object, the unknown speckle pattern, and the CTF of the microscope platform. We use a 40-μm defocus distance to effectively convert the complex amplitude information into intensity variations. A shorter distance may not be enough for the conversion. A longer distance, on the other hand, may lead to a reduction of the effective field of view. In our experiment, we first move the sample into the in-focus position using incoherent light illumination. We then move the sample to a 40-μm distance using a motorized stage. If the estimated defocus distance is not off too much, the CTF updating process can recover the distance numerically. Figure 6 shows the resolution gain of the reported platform using a USAF resolution target. Figure 6(a) shows the captured image under uniform illumination and Fig. 6(b) shows the raw near-field FP image under speckle illumination. Figure 6(c) shows the recovered object with different number of translated positions. The reconstruction quality increases as the number of captured raw images increases and we can resolve group 10 element 5 with a 310-nm linewidth. In our implementation, we can also recover the unknown speckle in a calibration experiment. The pre-recovered speckle pattern can then be used to recover other unknown objects. Figure 6(d) shows the recovered object with the pre-recovered speckle pattern. In this case, we can reduce the number of translated positions from ~1000 to ~300. Based on Figs. 6(c1) and 6(d1), we achieve a ~7.1-fold resolution gain in this experiment.

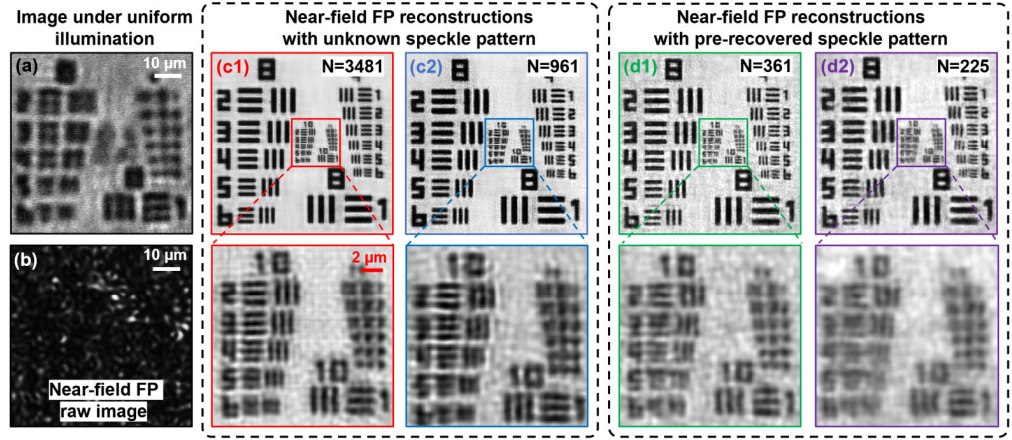

Fig. 6. Near-field FP for microscopy imaging. The captured raw image under uniform illumination (a) and speckle illumination (b). (c) The recovered object using different numbers of translated positions, with an unknown speckle pattern. (d) The recovered object using different numbers of translated positions and with a pre-recovered speckle pattern.

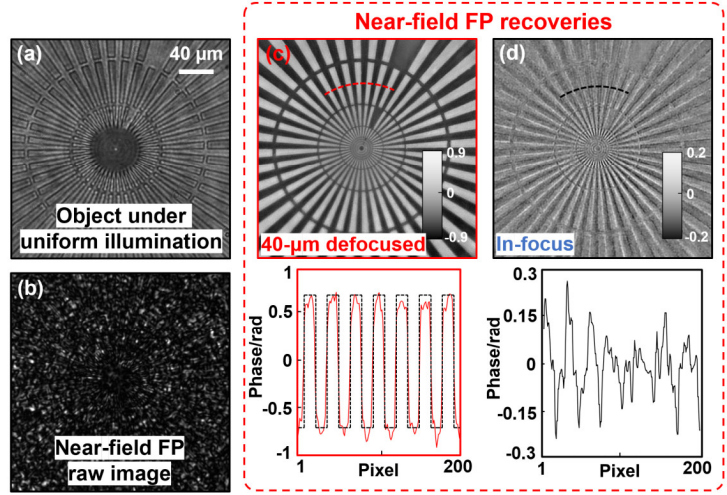

Fig. 7. Quantitative phase recovery via the near-field FP. The captured raw image under (a) uniform illumination and (b) speckle illumination. The recovered phase by placing the target at the 40-μm defocused distance (c) and the in-focus position (d). The phase profiles are plotted along the red and black dash arc in (c) and (d).

In the second experiment, we use a quantitative phase target (Benchmark QPT) as the object. Figure 7(a) shows the captured image under uniform illumination and Fig. 7(b) shows the near-field FP raw image under speckle illumination. Figure 7(c) shows the recovered phase by placing the object at the 40-μm defocus distance. As a comparison, Fig. 7(d) show the recovered phase for the in-focus position. We also plot the phase profile along the dash lines in 7(c) and 7(d). For the 40-μm defocus case, the recovered phase profile is in a good agreement with the theoretical phase value provided by the manufacturer (the dark dash line). This experiment validates the 'near-field' requirement of the proposed approach.

In the third experiment, we use a blood smear sample as the object. Figures 8(a) and 8(b) show the captured raw image under uniform and speckle illumination. The recovered intensity and phase of the blood smear are shown in Fig. 8(c), where the features of blood cells can be clearly resolved using our approach.

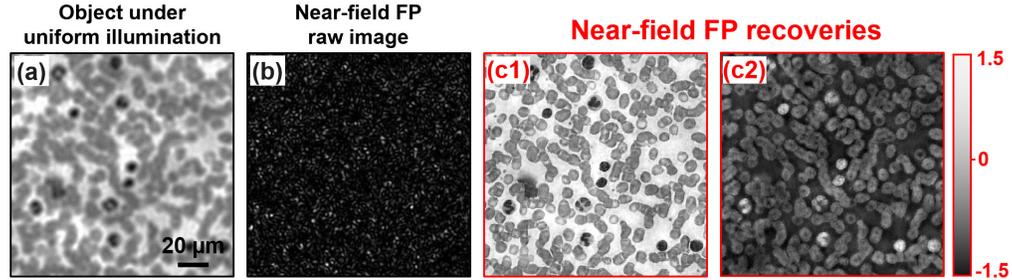

Fig. 8. Experimental demonstration of a blood smear sample using the reported approach. (a) The captured image of blood smear under uniform illumination (a) and speckle illumination (b). (c) The recovered intensity and phase using the near-field FP.

In many microscopy applications, it is important to achieve both high-resolution and wide field of view at the same time. We have also performed an experiment to demonstrate such a capability. In this experiment, we use a 2X, 0.1 NA objective lens to acquire images with a large field of view. Figure 9(a) shows the recovered image of the blood smear, with a size of 6.6 mm by 4.5 mm. Figure 9(b) shows the magnified views of Fig. 9(a), and Fig. 9(c) shows the corresponding raw images under uniform illumination.

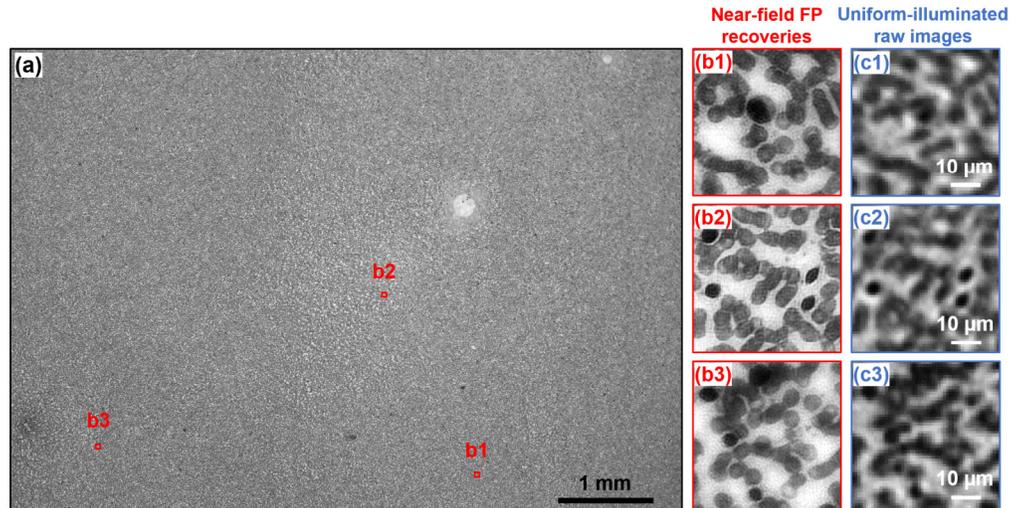

Fig. 9. Wide-field, high-resolution imaging via the near-field FP. (a) The recovered gigapixel intensity image of a blood smear section, with a field of view of 6.6 mm by 4.5 mm. (b1)-(b3) show three magnified view of (a). (c1)-(c3) The raw captured images under uniform illumination.

## 4. Near-field FP for long-range photographic imaging

The proposed near-field FP approach can be used for macroscale photographic imaging. The key innovation is to use a large-aperture lens to project speckle pattern with fine spatial features. We note that, aberration is not an issue for the large-aperture projection lens. The speckle feature size is only determined by the aperture size of the projection lens, regardless of the aberrations. Therefore, one can use a large plano-convex lens to project the fine speckle pattern.

Our experiment setup is shown in Fig. 10(a), where we use a large plano-convex lens (Thorlabs LA1050) and a diffuser (DG20-600) for speckle projection. On the detection side, we use a Nikon photographic lens (50 mm, f/1.8) with a circular aperture to acquire the images. The collection NA is ~0.0006 in our implementation. We then mechanically scan the speckle pattern to 41 by 41 different positions. The captured images are used to recover the high-resolution object image. Figure 10(b) shows the raw captured object image under uniform illumination. Figure 10(c) shows the raw near-field FP image under speckle illumination. The recovered object image is shown in Fig. 10(d), where we can resolve the 62-µm line width. Based on Figs. 10(b) and 10(d), we achieve ~7-fold resolution gain using this macroscale photographic imaging setup.

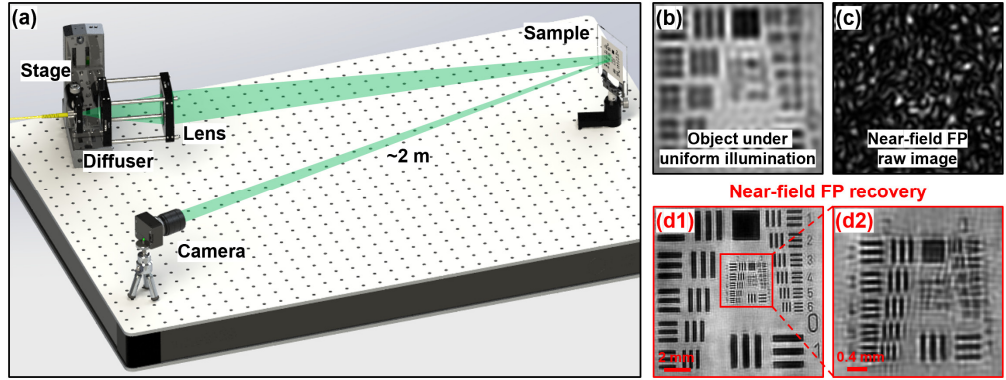

Fig. 10. Near-field FP for macroscale photographic imaging. (a) The experimental setup, where we use a diffuser and a large plano-convex lens to project the laser speckles onto the object. We use the same image-plane defocus distance as the microscope setup in Fig. 1(a). (b) The captured image of the USAF resolution target under uniform illumination. (c) The captured raw near-field FP image under speckle illumination. (d1) The high-resolution recovered object using the reported approach. (d2) The magnified view of the resolution target.

## 5. Summary

In summary, we have discussed a new coherent imaging modality, termed near-field Fourier ptychography, for tackling high-resolution imaging challenges in both microscopic and macroscopic imaging settings. Instead of directly resolving fine spatial features via high-resolution lenses, we project a speckle pattern with fine spatial features on the object. We then translate the object (or speckle) to different positions and acquire the corresponding images using a low-resolution lens. A Fourier ptychographic phase retrieval process is used to recover the complex object, the unknown speckle pattern, and the coherent transfer function at the same time. We achieve a ~7-fold resolution gain in both a microscopic and a macroscale photographic imaging setup. The final achievable resolution is not determined by the collection optics. Instead, it is determined by the feature size of the speckle pattern. Compared to FP, the reported approach can be readily employed in current synchrotron beamline setups and table-top transmission X-ray microscopes without major hardware modifications. It may find applications in light, coherent X-ray, and transmission electron imaging systems to increase resolution, correct aberrations, and provide quantitative absorption and phase contrast of the object.

## Appendix

The following simulation Matlab code consists of the following nine steps: 1) set the parameters for the coherent imaging system, 2) generate the input object image, 3) generate the CTF and the low-resolution captured image, 4) generate the speckle pattern, 5) generate the spiral sequence, 6) generate the low-resolution measurements, 7) reorder the measurement sequence for reconstruction, 8) generate the initial guess, 9) perform the iterative phase retrieval.

In step 1, we define the parameters for the imaging system, which are the same as our experimental setup.

%% Step 1: Set the parameters for the coherent imaging system
1. WaveLength = 0.53e-6; % Wave length (green)
2. PixelSize = 3.45e-6 /6.8; % Effective pixel size of imaging system
3. NA = 0.12; % Numerical aperture of imaging system
4. phasePatternIndex = 10*pi; % Maximum phase of pattern
5. phaseObjectIndex = 2*pi; % Maximum phase of object
6. InputImgSize = 257; % Size of image (assumed to be square)
7. DefocusZ = 40e-6; % Defocus distance
8. nIterative = 20; % Number of iterations
9. ShiftStepSize = 1; % Step size of each positional shift
10. SpiralRadius = 19;
11. nPattern = (SpiralRadius*2+1)^2; % Number of captured images

In step 2, we generate the complex input object, as shown in Fig. 11. The image's size is 257 by 257 pixels. The amplitude of the input object is normalized, and the input phase is set from -pi to pi. This complex object is stored in 'InputImg'.

%% Step 2: Generate the input object image
12. InputAmplitude = double(imread('fruits.png'));
13. InputAmplitude = imresize(InputAmplitude(:,:,1),[InputImgSize InputImgSize]);
14. InputAmplitude = InputAmplitude/max(InputAmplitude(:));
15. InputPhase = double(imread(('westconcordorthophoto.png')));
16. InputPhase = imresize(InputPhase,[InputImgSize,InputImgSize]);
17. InputPhase = phaseObjectIndex*(InputPhase/max(InputPhase(:))-0.5);
18. InputImg = InputAmplitude.*exp(1i.*InputPhase);
19. figure(1); subplot(1,2,1); imshow(abs(InputImg),[]); title('Ground Truth Amplitude');
20. subplot(1,2,2); imshow(angle(InputImg),[]); title('Ground Truth Phase');

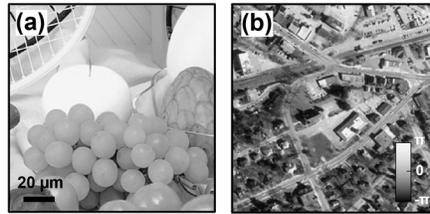

Fig. 11. The simulated high-resolution amplitude (a) and phase (b) of the input object.

In step 3, we generate the CTF of the imaging system. A phase pupil 'fmaskpro' is introduced to describe the effect of defocus aberration. Figures 12(a) and 12(b) show the CTF with no aberration (in-focus) and with a 40-μm defocus aberration, respectively. Figures 12(c) and 12(d) show the generated in-focus and defocused low-resolution images.

%% Step 3: Generate the CTF and the low-resolution image
21. k0 = 2*pi/WaveLength;
22. CutoffFreq = NA * k0;
23. kmax = pi/PixelSize;
24. [GridX, GridY] = meshgrid(-kmax:kmax/((InputImgSize-1)/2):kmax,-kmax:kmax/((InputImgSize-1)/2):kmax);
25. CTFInfocus = double((GridX.^2+GridY.^2)<CutoffFreq^2); % Pupil function, no aberration
26. LRInfocusTargetImgFT=fftshift(fft2(InputImg)).*CTFInfocus;

27. InputInfocusImgLR=abs(ifft2(ifftshift(LRInfocusTargetImgFT))).^2; % Low-resolution in-focus image
28. GridZ=sqrt(k0^2-GridX.^2-GridY.^2);
29. fmaskpro=exp(1j.*DefocusZ.*real(GridZ)).*exp(-abs(DefocusZ).*abs(imag(GridZ)));
30. CTFDefocus=fmaskpro.*CTFInfocus; % Defocus CTF
31. LRDefocusTargetImgFT=fftshift(fft2(InputImg)).*CTFDefocus;
32. InputDefocusImgLR=abs(ifft2(ifftshift(LRDefocusTargetImgFT))).^2; % Low-resolution defocus image
33. figure(2); subplot(2,2,1);imagesc(abs(CTFInfocus));title('infocus CTF');
34. subplot(2,2,2);imshow(InputInfocusImgLR,[]);title('LR infocus input image');
35. subplot(2,2,3);imagesc(angle(CTFDefocus));title('defocus CTF phase');
36. subplot(2,2,4);imshow(InputDefocusImgLR,[]);title('LR defocus input image');

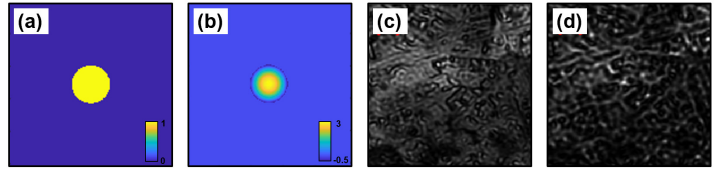

Fig. 12. The simulated CTF for low-resolution image generation. (a) The amplitude of simulated CTF. (b) The phase of simulated CTF with defocus aberration. The intensity of captured low-resolution image without aberration (c) and with the defocus aberration (d).

In step 4, we generate a speckle pattern with random phase ranged from 0 to 10*pi and random amplitude ranged from 0 to 1.
%% Step 4: Generate the speckle pattern
37. PatternAmplitude = rand(InputImgSize,InputImgSize);
38. PatternPhase = phasePatternIndex*rand(InputImgSize,InputImgSize);
39. Pattern = PatternAmplitude.*exp(1i.*PatternPhase);

In step 5, we generate a spiral sequence for object scanning. The object shifts along the 39-by-39-pixel spiral path. The x- and y- positions are stored in 'LocationX' and 'LocationY'.
%% Step 5: Generate the spiral sequence
40. LocationX = zeros(1,nPattern);
41. LocationY = zeros(1,nPattern);
42. SpiralPath = spiral(2*SpiralRadius+1);
43. for iShift = 1:nPattern
44.     [iRow, jCol] = find(SpiralPath == iShift);
45.     LocationX(1,iShift) = ShiftStepSize*(iRow-SpiralRadius);
46.     LocationY(1,iShift) = ShiftStepSize*(jCol-SpiralRadius);
47. end

In step 6, we generate the low-resolution image sequence by scanning the high-resolution object to different positions. We shift the object by multiplying an equivalent phase factor to the spectrum of the object, and the method is defined in the function 'subpixelshift()'. The low-resolution image sequence is obtained by applying the lowpass filter given by the CTF with the defocus aberration. The obtained low-resolution measurements with a dimension of 257 by 257 by 1521 pixels are stored in 'TargetImgs'. The first generated image is shown in Fig. 13(a).
%% Step 6: Generate the low-resolution measurements
48. TargetImgs = zeros(InputImgSize,InputImgSize,nPattern);
49. for iTargetImg = 1:nPattern
50.     ImgShiftTemp = subpixelshift(InputImg,LocationX(1,iTargetImg), LocationY(1,iTargetImg));
51.     ImgHighTemp = ImgShiftTemp.*Pattern;

```
52.     ImgHighFTTemp = fftshift(fft2(ImgHighTemp));
53.     ImgLowFTTemp = ImgHighFTTemp.*CTFDefocus;
54.     TargetImgs(:,:,iTargetImg) = abs(ifft2(ifftshift(ImgLowFTTemp))).^2;
55.     disp(iTargetImg);
56. end
57. figure(3); imshow(TargetImgs(:,:,1),[]); title('The 1st LR captured image');
58. function output_image = subpixelshift(input_image,xshift,yshift)
59.     [m,n,num] = size(input_image);
60.     output_image = (input_image);
61.     [FX,FY] = meshgrid(-floor(n/2):ceil(n/2-1),-floor(n/2):ceil(n/2-1));
62.     for i = 1:num
63.         Hs = exp(-1j*2*pi.*(FX.*xshift(1,i)/n+FY.*yshift(1,i)/m));
64.         output_image(:,:,i) = ifft2(ifftshift(fftshift(fft2(output_image(:,:,i))).*Hs));
65.     end
66. end
```

In step 7, we randomly reorder the generated low-resolution images for reconstruction. The order for reconstruction is stored in 'reOrder'.

```
%% Step 7: Reorder the image sequence for reconstruction
67. reOrder=randperm(nPattern);
68. TargetImgs_temp=TargetImgs(:,:,reOrder);
69. TargetImgs=TargetImgs_temp;
70. LocationX_temp = LocationX(reOrder); LocationY_temp = LocationY(reOrder);
71. LocationX = LocationX_temp; LocationY = LocationY_temp;
```

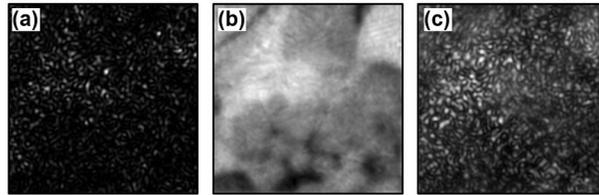

Fig. 13. The simulated captured low-resolution image sequence. (a) The first generated low-resolution measurement. The initial guess of the high-resolution object (b) and speckle (c).

In step 8, we generate the initial guess of the object and the speckle pattern, as shown in Figs. 13(b) and 13(c).

```
%% Step 8: Generate the initial guess of the object and the speckle pattern
72. ImgShifted = zeros(InputImgSize,InputImgSize,nPattern);
73. PatternShifted = zeros(InputImgSize,InputImgSize,nPattern);
74. for iTargetImg=1:nPattern
75.     ImgShifted(:,:,iTargetImg)=sqrt(subpixelshift(TargetImgs(:,:,iTargetImg),-
        LocationX(1,iTargetImg),-LocationY(1,iTargetImg)));
76.     PatternShifted(:,:,iTargetImg)=sqrt(TargetImgs(:,:,iTargetImg));
77.     disp(iTargetImg);
78. end
79. ImgMean = mean(ImgShifted,3); % Initial guess of the object
80. PatternMean = mean(PatternShifted,3); % Initial guess of the speckle pattern
81. figure(4); subplot(1,2,1); imshow(ImgMean,[]); title('Object initial guess');
82. subplot(1,2,2); imshow(PatternMean,[]); title('Speckle initial guess');
```

In step 9, we recover the object, speckle, and CTF using the iterative phase retrieval algorithm. The variable 'nIterative' defined in step 1 determines the number of iterations. In each iteration, the updated object, speckle, and CTF are stored in 'ImgRecovered',

'PatternRecovered', and 'CTFDefocus', respectively. The recovered high-resolution complex object and the speckle pattern are shown in Figs. 14(a)-14(c).

```
%% Step 9: Iterative reconstruction
83.  ImgRecovered = ImgMean;
84.  PatternRecovered = PatternMean;
85.  t=0; T=100; gama=0.2; eta=0.8; alpha=0.3;
86.  vImg=zeros(InputImgSize,InputImgSize);
87.  vP=zeros(InputImgSize,InputImgSize);
88.  ImgRecoverBeforeUpdate=ImgRecovered;
89.  PatternRecoverBeforeUpdate=PatternRecovered;
90.  for iterative = 1:nIterative
91.  for iTargetImg = 1:nPattern
92.  t=t+1;
93.  ImgRecoveredTemp = subpixelshift(ImgRecovered,LocationX(1,iTargetImg),
         LocationY(1,iTargetImg));
94.  TempTargetImg = PatternRecovered.*ImgRecoveredTemp;
95.  TempTargetImgCopy = TempTargetImg;
96.  TempTargetImgFT = fftshift(fft2(TempTargetImg));
97.  LRTempTargetImgFT = CTFDefocus.*TempTargetImgFT;
98.  LRTempTargetImg = ifft2(ifftshift(CTFDefocus.*TempTargetImgFT
99.  LRTempTargetImg_AmpUpdated =
         sqrt(TargetImgs(:,:,iTargetImg)).*exp(1i.*angle(LRTempTargetImg)); % Update
         the amplitude and keep the phase unchanged
100. LRTempTargetImg_AmpUpdatedFT = fftshift(fft2(LRTempTargetImg_AmpUpdated));
101. TempTargetImgFT =
         TempTargetImgFT+conj(CTFDefocus)./(max(max((abs(CTFDefocus)).^2)))
         .*(LRTempTargetImg_AmpUpdatedFT-LRTempTargetImgFT);
102. CTFDefocus=CTFDefocus+gama.*conj(TempTargetImgFT)./(max(max((abs(TempTarget
     ImgFT)).^2)))  .*(LRTempTargetImg_AmpUpdatedFT-LRTempTargetImgFT); % Update
     the defocus CTF
103. TempTargetImg = ifft2(ifftshift(TempTargetImgFT));
104. PatternRecovered =
         PatternRecovered+gama.*conj(ImgRecoveredTemp).*(TempTargetImg-
         TempTargetImgCopy)./((1-alpha).*(abs(ImgRecoveredTemp)).^2
         +alpha.*max(max(abs(ImgRecoveredTemp).^2))); % Update the speckle
105. ImgRecoveredTemp =
         ImgRecoveredTemp+gama.*conj(PatternRecovered).*(TempTargetImg-
         TempTargetImgCopy)./((1-alpha).*(abs(PatternRecovered)).^2
         +alpha.*max(max(abs(PatternRecovered).^2)));  % Update the object
106. ImgRecovered = subpixelshift(ImgRecoveredTemp,-LocationX(1,iTargetImg),
         -LocationY(1,iTargetImg));
107. ImgRecoverAfterUpdate= ImgRecovered;
108. PatternRecoverAfterUpdate=PatternRecovered;
     % Add momentum to accelerate convergence
109. if t==T
110.    vImg=eta*vImg+(ImgRecoverAfterUpdate-ImgRecoverBeforeUpdate);
111.    ImgRecovered=ImgRecoverAfterUpdate+eta*vImg;
112.    ImgRecoverBeforeUpdate=ImgRecovered;
113.    vP=eta*vP+(PatternRecoverAfterUpdate-PatternRecoverBeforeUpdate);
114.    PatternRecovered= PatternRecoverAfterUpdate+eta*vP;
115.    PatternRecoverBeforeUpdate=PatternRecovered;
116.    t=0;
```

```
117. end
118. disp([iterative iTargetImg]);
119. end
120. figure(5); set(gcf,'outerposition',get(0,'screensize'));
121. subplot(1,3,1); imshow(abs(ImgRecovered),[]); title('Recovered object amplitude');
122. subplot(1,3,2); imshow(angle(ImgRecovered),[]); title('Recovered object phase');
123. subplot(1,3,3); imshow(abs(PatternRecovered),[]); title('Recovered pattern amplitude');
124. pause(0.01);
125. end
```

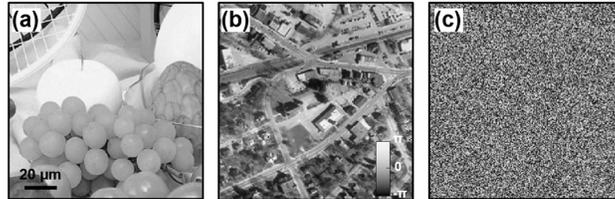

Fig. 14. The recovered high-resolution amplitude (a), phase (b), and the speckle (c).

## Funding

National Science Foundation (1510077); National Institute of Health (R21EB022378, R03EB022144). H. Zhang acknowledges the support of the China Scholarship Council.